\documentclass[prd,amsfonts,twocolumn,nofootinbib,showpacs]{revtex4}
\RequirePackage{graphicx}
\usepackage{enumerate}

\begin{document}

\title{Curvaton mechanism after multi-field inflation}

\author{Seishi Enomoto}
\affiliation{Department of Physics, Nagoya University, Nagoya 464-8602, Japan}
\author{Tomohiro Matsuda}
\affiliation{Department of Physics, Lancaster University,  
Lancaster LA1 4YB, UK, and
 Laboratory of Physics, Saitama Institute of Technology,
Fukaya, Saitama 369-0293, Japan}

\begin{abstract}
The evolution of the curvature perturbation after multi-field
inflation is studied in the light of the curvaton mechanism.
Past numerical studies show that many-field inflation causes
 significant evolution of the curvature perturbation after inflation,
 which generates significant non-Gaussianity at the same time.  
We reveal the underlying mechanism of the evolution and show 
that the evolution is possible in a typical two-field inflation model.
\end{abstract}

\maketitle

\section{Introduction}

The  primordial curvature perturbation $\zeta(k)$
is strongly constrained by observation and provides a unique window
on the very early universe~\cite{Lyth-book}. 
It is known to have the spectrum ${\cal P}_\zeta(k)\simeq (5\times
10^{-5})^2$ with spectral tilt
$n-1\equiv d\ln {\cal P}_\zeta/d\ln k \simeq -0.04$, and in future one
could detect the running $dn/d\ln k$ as 
well as non-Gaussianity signaled by the bispectrum and trispectrum.

The process of generating $\zeta$ begins
presumably during inflation, when the vacuum fluctuations of one or more
bosonic fields are converted to classical perturbations. Within this general
framework, there exist many proposals~\cite{Lyth-book}.

One proposal is to use two or more inflaton fields, which drive
inflation in the multi-field model. 
That paradigm has been widely investigated, but
it has usually been supposed that $\zeta(x,t)$ evaluated at an epoch
$t_\mathrm{end}$ just before
(or sometimes just after)  the end of inflation
is to be identified with the observed quantities in the spectrum.
For this reason, a great deal of effort has gone into the calculation
of the spectrum, bispectrum and trispectrum of $\zeta$
at the end of inflation~\cite{Multi1,Multi2,Multi3,Multi4,Multi5,Multi6,
Multi-matsuda, Multi-NG}.

The evolution after many-field inflation has been studied numerically
in Ref.~\cite{CGJ} using the statistical distribution of the
parameters~\cite{Easther:2005zr, Battefeld:2008bu}. 
Later in Ref.~\cite{afterCGJ} the evolution of
the non-Gaussianity has been investigated.
In these studies it has been found that there is a minimal number of the
inflaton field $N_f\ge 10^3$, which is needed to realize  
the late-time creation and the domination of the curvature perturbation.
Also, the number $N_f$ has been related to the creation of the 
non-Gaussianity.
On the other hand, the calculation is not analytic and it is not clear
if the evolution is possible in a two (or a few)-field model.

In this paper, we point out that the actual calculation of the curvature
perturbation might well depend on the evolution after multi-field
inflation, even if the number $N_f$ is {\em not large}.
We show that the minimum number is $N_f=2$, simply because the
mechanism requires isocurvature perturbation.

Just for simplicity, consider $N_f=2$ with the light scalar fields
($\phi, \sigma$) during inflation. 
The adiabatic and the entropy directions of multi-field inflation 
are defined using those fields.  
Basically, the ``inflaton'' (the adiabatic field) is not
identical to $\phi$, even if $\sigma$ plays the role of the curvaton.
The mixing is negligible when $\sigma$ is much
lighter than $\phi$; that is the limit where the usual curvaton scenario
applies.

Alternatively, it is possible to consider the opposite limit, where the
fields have nearly equal mass\footnote{In
Ref.~\cite{CGJ, Easther:2005zr, Battefeld:2008bu, afterCGJ}, 
statistical distribution of the inflaton mass
has been considered for
N-flation. 
The deviation $m_\mathrm{Max}/m_\mathrm{min}\lesssim O(10)$
will be considered in this paper.}.
Can the curvaton mechanism work in that limit?  
A naive speculation is that the biased initial condition 
($\sigma/\phi\ll 1$) might lead to the curvaton
mechanism in that limit.
Indeed the speculation is correct; however
to reach the correct conclusion we need quantitative calculation of the curvaton
mechanism in the equal-mass limit.
The calculation details are shown in the Appendix.
The usual curvaton mechanism is reviewed in Sec.\ref{sec:prepare}, and
the non-linear formalism of the curvaton mechanism is reviewed in
Sec.\ref{sec:3}.
The basic idea of the equal-mass curvaton model is shown in
Sec.\ref{sec:4} for two-field inflation.
Deviation from the equal-mass limit and the applications are discussed
in Sec.\ref{sec:5}.

\section{Curvaton mechanism}
\label{sec:prepare}
In this section we review $\delta N$ formalism used to calculate
$\zeta$. To define $\zeta$ one smooths the energy density $\rho$ 
on a super-horizon scale shorter than any scale of interest. Then it satisfies
the local energy continuity equation,
\begin{equation}
\frac{\partial \rho(x,t) }{\partial t} = - \frac{3}{a(x,t)}
\frac{\partial a(x,t)}
{\partial t} \left( \rho(x,t) + p(x,t) \right)
, \end{equation}
where $t$ is time along a comoving thread of spacetime and $a$ is the local
scale factor. Choosing the slicing of uniform $\rho$, the curvature 
perturbation is $\zeta\equiv \delta (\ln a)$ and
\begin{equation}
\frac{\partial \zeta (x,t) }{\partial t} = \delta \left( \frac{\dot\rho(t)}
{\rho(t) + p(x,t) } \right)
.\end{equation}
If $p$ is a function purely of $\rho$, one will find $\dot\zeta=0$. 
That is the case of single field inflation when no other field
perturbation is relevant.
The inflaton field $\phi(x,t)$  determines the future evolution 
of both $\rho$ and $p$.
Similarly, the component perturbations $\zeta_i$
are conserved if they scale
like matter ($\rho_m\propto a^{-3}$) or radiation ($\rho_r\propto a^{-4}$).

During nearly exponential inflation, the vacuum fluctuation of 
each light scalar field $\phi_i$ 
is converted at horizon exit to a nearly Gaussian
classical perturbation with spectrum $(H/2\pi)^2$, where 
$H\equiv \dot a(t)/a(t)$ in the unperturbed universe. Writing 
\begin{equation}
\zeta = \delta [ \ln (a(x,t)/a(t_1)] \equiv \delta N
, \end{equation}
 and taking $t_*$ to be an epoch 
during inflation after relevant scales leave the horizon,
we define $N(\phi_1(x,t_*),\phi_2(x,t_*),\cdots,t,t_*)$ so that 
\begin{equation}
\zeta(x,t) = N_i \delta \phi_i(x,t_*)
+ \frac{1}{2} N_{ij} \delta \phi_i(x,t_*)\delta \phi_j(x,t_*) + \cdots
, \end{equation}
where a subscript $i$ denotes $\partial/\partial \phi_i$ evaluated on the 
unperturbed trajectory.
We find
\begin{eqnarray}
n-1 &=&  \frac{2\sum_i N_i N_j \eta_{ij}}{\sum_m N_m^2} 
-2\epsilon - \frac{2}{M_p^2 \sum_m N_m^2} \\
\eta_{ij} &\equiv& M_p^2V_{ij}/V,\qquad \epsilon \equiv M_p^2
\sum_m V_m^2/V^2, 
\end{eqnarray}
 where $M_p$ is the reduced Planck mass.

The standard curvaton model~\cite{lm, curvaton-paper} 
assumes that these expressions are dominated by the single `curvaton' field
$\sigma$, which starts to oscillate
during radiation domination at a time 
when the component perturbation $\zeta_\sigma$ has negligible
contribution to the curvature perturbation.
Then the non-Gaussianity parameter is given by~\cite{Lyth-gfc, Lyth-general}
\begin{eqnarray}
f_{NL} &\simeq&  \frac{5}{4r_\sigma} \left( 1 +\frac{g''g}{g^2}\right)
-\frac{5}{3} - \frac{5}{6}r_\sigma,
\end{eqnarray}
where $g(\sigma)$ is the initial amplitude of the oscillation as a function
of the curvaton field at horizon exit~\cite{Lyth-gfc}.
Here $r_\sigma$ is identical to $r_1$, which will be defined in this
paper\footnote{In this paper
we use $(\phi_1, \phi_2)$ for two-field inflation, instead of
using the conventional $(\sigma, \phi)$ in the curvaton scenario.}.

\section{Non-linear formalism and the evolution of the perturbation}
\label{sec:3}
In this paper we consider a clear separation of the adiabatic and
the entropy perturbations in a two-field inflation model.
The non-linear formalism for the component curvature perturbation is defined in
Ref.~\cite{Lyth-general, Langlois:2008vk} as 
\begin{eqnarray}
\label{def-compzeta}
\zeta_i&=&\delta N+\int^{\rho}_{\bar{\rho}_i}H
\frac{d\tilde{\rho}_i}{3(1+w_i)\tilde{\rho}_i}\nonumber\\
&=& \label{ln-rho}\delta N + \frac{1}{3(1+w_i)}\ln
 \left(\frac{\rho_i}{\bar{\rho}_i}\right)\nonumber\\
&\simeq & \delta N+\frac{1}{3(1+w_i)}\frac{\delta \rho_i^\mathrm{iso}}{\bar{\rho}_i},
\end{eqnarray}
where $w_i=1/3$ for the radiation fluid  and $w_i=0$ for the matter fluid.
Here a bar is for a homogeneous quantity,  and the curvature perturbation
of the total fluid should be
discriminated from the component curvature perturbation $\zeta_i$.
The quantity $\delta \rho_i^\mathrm{iso} =\rho_i-\bar{\rho}_i$ in
Eq.(\ref{def-compzeta}) is the
isocurvature perturbation (the fraction perturbation that satisfies $\sum \delta
\rho_i^\mathrm{iso}\equiv 0$), 
which is defined on the uniform density hypersurfaces.

In order to formulate the evolution of the curvature 
perturbation, which is caused by the adiabatic-isocurvature mixings,
we need to define first the ``starting point'' perturbations at an
epoch.

\subsection{The primordial perturbations}
For the first step, we define the primordial quantities.
In this paper the quantities at the end of inflation are denoted by the
subscript ``end'', while the corresponding scale exited horizon at $t_*$.
The subscript ``$*$'' is used for the quantities at the horizon exit.
For our purpose, we define the primordial curvature and
isocurvature perturbations at the end of the primordial inflation.

We find from Eq.(\ref{ln-rho}); 
\begin{eqnarray}
\rho_i&=&\bar{\rho}_ie^{3(1+w_i)(\zeta_i-\delta N)}\nonumber\\
&\simeq& \bar{\rho}_i+3(1+w_i)\zeta_i^\mathrm{iso}\bar{\rho}_i\nonumber\\
&\equiv&  \bar{\rho}_i+\delta \rho_i^\mathrm{iso}.
\end{eqnarray}
Then we find from
$\rho^\mathrm{tot}\equiv\rho_1+\rho_2=\bar{\rho}_1+\bar{\rho}_2$:
\begin{equation}
\label{trivial-iso}
f_1e^{3(1+w_1)(\zeta_1-\delta N)}
+\left(1-f_1 \right)e^{3(1+w_2)(\zeta_2-\delta N)}=1,
\end{equation}
where the fraction of the energy density is defined by
\begin{equation}
f_1 \equiv \frac{\bar{\rho}_1}{\bar{\rho}_1+\bar{\rho}_2}.
\end{equation}
Expanding Eq.(\ref{trivial-iso}) and solving the equation for $\delta
N$,  we find at first order~\cite{Lyth-general}
\begin{eqnarray}
\label{deltaN-1}
\delta N&=&r_1 \zeta_1+(1-r_1)\zeta_2\nonumber\\
&\equiv& \left[r_1\zeta_1^\mathrm{iso}+(1-r_1)\zeta_2^\mathrm{iso}\right]
+\zeta^\mathrm{adi},
\end{eqnarray}
where $\zeta^\mathrm{iso}_i$ denotes the second component in
Eq.(\ref{def-compzeta}).
$r_1$ is defined by
\begin{equation}
\label{r1-basic}
r_1\equiv
\frac{3(1+w_1)\bar{\rho}_1}{3(1+w_1)\bar{\rho}_1+3(1+w_2)\bar{\rho}_2}.
\end{equation}

Defining the primordial adiabatic curvature perturbation
($\zeta^\mathrm{inf}$) just at the end of inflation, 
the component curvature perturbation ($\zeta_i$) can be split into
$\zeta^\mathrm{inf}$ and $\zeta_i^\mathrm{iso}$.

The obvious identity is
\begin{equation}
\label{iso0}
r_{1,\mathrm{end}}\zeta_{1,\mathrm{end}}^\mathrm{iso}+(1-r_{1,\mathrm{end}})\zeta_{2,\mathrm{end}}^\mathrm{iso}
\equiv  0,
\end{equation}
which is valid at the end of inflation.
Apart from that point the deviation due to the evolution of
$r_1$ becomes significant.

The parameter of the fluid ($w_i$) is constant when $\rho_i$ behaves like
matter ($w_i=0$) or radiation ($w_i=1/3$), and
a jump (e.g, $w_i=0 \rightarrow w_i=1/3$) is possible when instant
transition is assumed.
In this paper we are using the sudden-decay approximation for the
curvaton mechanism.~\footnote{Authors of Ref.~\cite{Enqvist:2011jf}
consistently accounted the curvaton decay when the curvaton decays
pertubatively and showed that the correction to the potential can be
significant.}
We also assume that the inflatons start sinusoidal oscillations just at
 the end of slow-roll.

The curvature perturbation in the standard
curvaton scenario is usually expressed as
\begin{equation}
\label{lang-eq}
\delta N =r_1\zeta_1+(1-r_1)\zeta^\mathrm{inf}.
\end{equation}
Assuming that $\zeta_1^\mathrm{iso} \gg  \zeta^\mathrm{inf}  \gg
\zeta_2^\mathrm{iso}$, one will find
$\zeta_1 \simeq  \zeta_1^\mathrm{iso}$ and $\zeta_2\simeq
\zeta^\mathrm{inf}$, which gives Eq.(\ref{lang-eq}) from
Eq.(\ref{deltaN-1}). 
Usually the above approximation is justified 
when $m_1\ll m_2$ and the curvaton is negligible during
inflation. 

In this paper we are considering the equal-mass limit ($m_1\simeq m_2$),
which is in the opposite limit of the conventional curvaton.
In Appendix we show the validity of the above approximations and
derive the quantitative bound on the ratio $\phi_1/\phi_2$.

\section{A basic model}
\label{sec:4}
In this section we show why 
the curvaton mechanism can create the dominant part of the curvature
perturbation after conventional chaotic multi-field inflation,
 neither by adding extra light field (curvaton) nor by
introducing many inflatons. 
The calculation clearly explains why and how the curvaton mechanism works 
in the equal-mass limit ($m_1\simeq m_2$).

We assume (for simplicity) that after inflation the field $\phi_2$
decays immediately into radiation and $\phi_1$ starts
sinusoidal oscillation at the same time. 
Then $\phi_1$ decays late at $H_{d1}\ll H_I$. 
There is no mixing between these components.
Here $H_I$ denotes the Hubble parameter during primordial inflation. 

In this scenario, we consider two phases $(A,B)$ characterized by 
$w_{1A}=0$ and $w_{1B}=1/3$.
Here the subscripts $A$ and $B$ denotes the quantities in the phase A
and B.
They are separated by the uniform density
hypersurface $H_{d1}\simeq \Gamma_1$:
\begin{enumerate}[(A)]
\item $\rho_1$; oscillation, $\rho_2$; radiation\\
 ($w_1 = 0$, $w_2 =1/3$)
\item Radiation\\
  ($w_1 =w_2 =1/3$).
\end{enumerate}

The important assumption of the model is that the transition occurs on the
 uniform density hypersurfaces so that we can neglect additional creation of 
$\delta N$ (modulation) at the transition.

We find in phase (A); 
\begin{eqnarray}
\label{deltanbc}
\delta N &\equiv& r_{1A} \zeta_{1A} + (1-r_{1A})\zeta_{2},
\end{eqnarray}
where the subscript ``$A$'' (or ``$B$'') is omitted for $\zeta_{2}$,
since $\zeta_2$ is constant during the evolution.
Here we used the definition
\begin{equation}
\label{rminus}
r_{1A} =\frac{3\bar{\rho}_1}{3\bar{\rho}_1+4\bar{\rho}_2}.
\end{equation}

\begin{figure}[b]
\centering
\includegraphics[width=1.0\columnwidth]{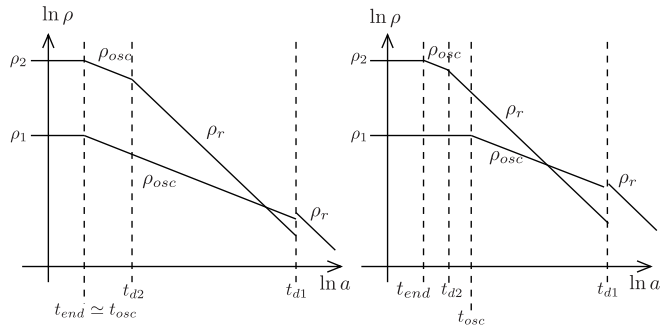}
 \caption{$t_\mathrm{end}$, $t_{d2}$, $t_\mathrm{osc}$
 and $t_{d1}$ denote the time at the end of inflation, $\phi_2$
 decay, the beginning of 
 $\phi_1$ oscillation and $\phi_1$ decay, respectively.
Our scenario is shown in the left-hand side, which gives
the time-ordering $t_{end}\simeq t_{osc}<t_{d2}<t_{d1}$.
The usual curvaton scenario is shown in the right-hand side, which 
gives $t_{end}<t_{d2}<t_{osc}<t_{d1}$.}
\label{fig:figure1}
\end{figure}

Consider a simple double-quadratic chaotic inflation model in the equal-mass
limit.
The potential is given by
\begin{equation}
\label{sym-pot}
V(\phi_1,\phi_2)=\frac{1}{2}m^2\left(\phi_1^2+\phi_2^2\right)\equiv
\frac{1}{2}m^2 \phi_r^2,
\end{equation}
where $\phi_{1,2}$ are real scalar fields.
Besides the potential, we need the interaction that causes difference in
the decay rates. 
Fig.\ref{fig:figure1} shows the evolution of the densities after
 inflation.
The end of chaotic inflation is given by
\begin{equation}
\phi_{1,\mathrm{end}}^2+\phi_{2,\mathrm{end}}^2\equiv 
\phi_{r,\mathrm{end}}^2\simeq M_p^2.
\end{equation}
\begin{figure}[tb]
\centering
\includegraphics[width=0.5\columnwidth]{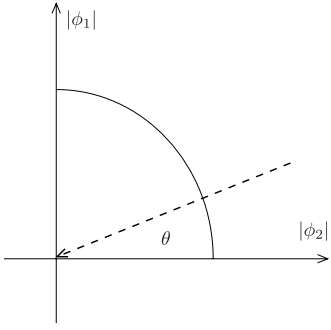}
 \caption{The straight dotted line with an arrow is the inflaton
 trajectory, and the circle gives the uniform-density surface along which the entropy
  perturbation $\delta s$ appears.}     
\label{fig:equalmass}
\end{figure}
Since the potential is quadratic during inflation, we
find
\begin{equation}
\zeta^\mathrm{inf}=\frac{1}{\eta}\frac{\delta \phi_{r*}}{\phi_{r*}}.
\end{equation}
In this section we consider $\theta\ll 1$, which leads to
the simplifications $\sin\theta\sim \theta$ and $\cos\theta\sim 1$.
Our approximations are based on the exact calculation
in Appendix A.

From Eq.(\ref{zeta12ndorder}), we find the component perturbation of
the late-decaying component ($\phi_1$) at the end of inflation:
\begin{equation}
\zeta_{1A} \simeq \frac{1}{3}\frac{\delta
 \rho_{1,\mathrm{end}}^\mathrm{iso}}{\bar{\rho}_{1,\mathrm{end}}} \simeq 
\frac{2}{3}\frac{\delta\theta}{\bar{\theta}}+
\frac{1}{3}\left(\frac{\delta\theta}{\bar{\theta}}\right)^2.
\end{equation}
The usual approximation of the curvaton mechanism is
$\zeta_{1A}^\mathrm{iso} \gg  \zeta^\mathrm{inf}$.
The validity of this approximation is examined in the Appendix.

From Eq.(\ref{zeta-losecond}), the final curvature perturbation is 
\begin{equation}
\zeta^\mathrm{fin}\simeq \frac{2r_{1-}}{3}
\left[\frac{\delta \theta}{\bar{\theta}}+\frac{1}{2}\left(\frac{\delta
				    \theta}{\bar{\theta}}\right)^2
\right].
\end{equation}
Defining the ratio $y\equiv \sqrt{\Gamma_1/\Gamma_2}$, 
$r_{1-}$ ($r_{1}$ evaluated in the phase (A) just before the decay) is given by
\begin{equation}
r_{1-}\simeq \frac{3\bar{\theta}^2}{3\bar{\theta}^2+4y}.
\end{equation}

The non-Gaussianity parameter has been calculated in Ref.~\cite{Lyth-general}.
We find for $\theta\ll 1$:
\begin{eqnarray}
f_{NL} &\simeq&  \frac{5}{4r_1} \left( 1 +\frac{g''g}{g^2}\right)
-\frac{5}{3} - \frac{5}{6}r_1\nonumber\\
&\sim&\frac{5}{4r_{1-}}.
\end{eqnarray}

Further simplification is possible when $\delta \theta=\delta
s_*/\phi_{r*}$ and ${\cal P}_{\delta s_*}={\cal P}_{\delta \phi_{r*}}$. 
For the quadratic potential we have $\phi_{r*}=2\sqrt{N_e}M_p \gg
\phi_{e,\mathrm{end}}$, where 
$N_e$ is the number of e-foldings  during the primordial inflation
spent after the corresponding scale exits horizon.
The condition of the curvaton mechanism $\zeta^\mathrm{fin}>\zeta^\mathrm{inf}$ gives 
\begin{equation}
\label{barthup}
\bar{\theta}<\frac{2}{3}r_{1-}\eta\simeq \frac{5}{6}\frac{\eta}{f_{NL}}.
\end{equation}
Here $\bar{\theta}$ should be less than 1 but does not require many orders of
magnitude. 
From the CMB spectrum we find the normalization given by
\begin{equation}
{\cal P}^{1/2}_{\zeta^\mathrm{fin}}\simeq
 \frac{r_{1-}}{6\pi\sqrt{N_e}\bar{\theta}}\frac{H_I}{M_p}\simeq 5\times
 10^{-5}.
\end{equation}
Using Eq.(\ref{barthup}), we find
\begin{equation}
\frac{H_I}{M_p}< 5\eta  \times 10^{-3},
\end{equation}
which does not always require significant suppression.
The ratio  $y\equiv \sqrt{\Gamma_1/\Gamma_2}$
  is calculated in Eq.(\ref{appy}) and is given by
\begin{equation}
y\simeq  \frac{3}{5}f_{NL}\theta^2.
\end{equation}
We thus find that the difference between $\phi_1$ and $\phi_2$ decay rates is
in the conceivable range. 

The above conditions tell us how small $\theta$ and $y$ have to be to
get a given CMB spectrum and $f_{NL}$.
They have to be some orders of magnitude below 1 but not very many.

If the potential during inflation is both symmetric
and quadratic, we find $\eta\equiv \eta_1=\eta_2$.
We thus find the spectral index
\begin{equation}
n-1=-2\epsilon+\eta= 0,
\end{equation}
which shows that the above model requires deviation
from the symmetric potential.

Looking back into the many-field inflation, the model in Ref.~\cite{CGJ}
assumed that the inflaton masses are not exactly the same but may have
statistical distribution around the mean value.
In that case, the cancellation in the spectral index is not realistic.
Since the deviation from the symmetric potential is expected, we need to 
examine what deviation is needed for the model.
Then we can understand why and how the curvaton mechanism 
works in the many-field inflation model.

\section{Deviation from the symmetric potential}
\label{sec:5}
The deviation from the symmetric quadratic potential  can be
classified as follows;
\begin{enumerate}
\item \underline{A small mass difference ($1\lesssim m_2/m_1 \lesssim 10$)}

The spectral index does not vanish when the double quadratic potential
has different (but not so much different as the usual curvaton) mass.
The slow-roll parameters are
\begin{eqnarray}
\epsilon_H &\equiv& \frac{\dot{H}}{H^2}= \sum \epsilon_i
=\sum\eta_i f_i
\nonumber\\
\eta_i&\equiv& \frac{m_i^2}{3H_I^2},
\end{eqnarray}
where the fraction of the density is given by $f_i\equiv
      \frac{\rho_{i*}}{\rho_\mathrm{tot*}}$.
The spectral index is shifted from $n_s-1=0$ and is given by
\begin{eqnarray}
\label{spect-1}
n_s-1&=&-2\epsilon_H+2\eta_1\nonumber\\
&\simeq& -2[\eta_1 f_1+\eta_2(1-f_1)]+2\eta_1\nonumber\\
&\simeq& -2(\eta_2-\eta_1)\nonumber\\
&\equiv& -2P\eta_2\nonumber\\
&=&-\frac{P}{N_e},
\end{eqnarray}
where $P\equiv \frac{m_2^2-m_1^2}{m_2^2}<1$.
The observation~\cite{WMAP7} shows $n_s-1=0.037\pm 0.014$,
which suggests $N_e \lesssim 40$ and 
requires secondary inflation~\cite{thermal-Inf}.

Besides the spectral index, $m_1<m_2$ suggests that the oscillation of
the field $\phi_1$ is slightly delayed compared to $\phi_2$.
The delay may enhance the density of $\phi_1$ at the beginning of the
      oscillation, while the initial $\rho_1$ 
      density may be reduced since $m_1$ is smaller.
Defining $y_\mathrm{eff}\equiv
\sqrt{\Gamma_1/m_1}$ and $\theta_\mathrm{eff} \equiv \phi_1/\phi_2$ at
      the end of inflation, we find
\begin{eqnarray}
r_{1A-}&\simeq&\frac{3m_1^2 \bar{\phi}_1^2}{3m_1^2 \bar{\phi}_1^2+4m_2^2\bar{\phi}_2^2
 \left(\frac{m_1^2}{m_2^2}\right)y_\mathrm{eff}}\nonumber\\
&\simeq&\frac{3\bar{\theta}_\mathrm{eff}^2 }{3\bar{\theta}_\mathrm{eff}^2 +4 y_\mathrm{eff}},
\end{eqnarray}
which gives a similar bound for $\theta_\mathrm{eff}$ ($y_\mathrm{eff}$).

\item \underline{Heavy curvaton ($m_1\gtrsim m_2$)}

Usually the curvaton is assumed to be much lighter than 
the inflaton; however this assumption could be avoided.
We consider the curvaton mechanism when
the curvaton is slightly heavier than the inflaton.

We assume $\rho_2>\rho_1$. 
Once it is assumed at the beginning of inflation, it remains true during
      inflation.\footnote{The opposite condition ($\rho_1>\rho_2$) requires
      $\phi_{1*}>M_p$, which suppresses the component perturbation of
      the curvaton and does not realize the curvaton mechanism.} 
Then $\phi_1$-oscillation starts {\em during inflation}.
It begins when
\begin{equation}
m_1^2=H^2_\mathrm{osc}\simeq\frac{m_2^2\phi_2^2|_\mathrm{osc}}{6M_p^2},
\end{equation}
where the subscript ``osc'' denotes the beginning of $\phi_1$-oscillation.
From the above equation and $\phi_2|_\mathrm{osc}\simeq2\sqrt{N_2}M_p$, where $N_2$ is the
remaining number of e-foldings after the beginning of $\phi_1$-oscillation, we find
\begin{equation}
\label{n2-mm}
N_2=\frac{3 m_1^2}{2m_2^2}.
\end{equation}
Defining $y_\mathrm{eff}\equiv e^{3N_2}\sqrt{\Gamma_1/\Gamma_2}$ and $\theta_\mathrm{osc}\equiv
      [\phi_1/\phi_2]_\mathrm{osc}$, we can estimate
\begin{eqnarray}
r_{1A-}
&\sim&\frac{3\bar{\theta}_\mathrm{osc}^2
 }{3\bar{\theta}_\mathrm{osc}^2 +4 y_\mathrm{eff}}.
\end{eqnarray}
Unfortunately, the spectral index is
\begin{eqnarray}
n_s-1
&\simeq& -2\epsilon_H+2\eta_1\nonumber\\
&\simeq &-2\eta_2+2\eta_1\nonumber\\
&\simeq& 2\eta_1>0.
\end{eqnarray}

\item \underline{Symmetric but Non-quadratic}

The potential could be dominated by a polynomial 
$V(\phi_r)\propto\phi_r^p$ at the moment when the
      perturbation exits horizon, while it can be
      approximated by the quadratic potential during the oscillation.
For the polynomial we find $\phi_{r*}\simeq \sqrt{2pN_e}M_p$ and
      the slow-roll parameters
\begin{eqnarray}
\epsilon_H &\simeq& \frac{1}{2}M_p^2\frac{p^2}{\phi_r^2} \\
\eta_1&\simeq& M_p^2\frac{p(p-1)}{\phi_r^2}.
\end{eqnarray}
The spectral index is shifted and is given by
\begin{eqnarray}
n_s-1&\simeq& - \frac{M_p^2}{\phi_r^2}\left[p^2-2p(p-1)\right]\nonumber\\
&\simeq& \frac{p-2}{2N_e}.
\end{eqnarray}
The result suggests that $p<2$ is needed for the scenario.
In that case the mass and the coefficient of the polynomial must run
in the trans-Planckian~\cite{run-inf}.
$p=1$ would correspond to monodromy in the string theory and it requires 
$N_e\lesssim 20$.
$p<1$ is an interesting possibility if the effective action allows
fractional power.

\end{enumerate}

\subsection{A model with a complex scalar}
An inderesting application of the idea is that a conventional 2-field multiplet
contains both inflation and the curvaton at the same time.
Consider a complex scalar field $\Phi\equiv \phi_2+i\phi_1$,
which gives the symmetric potential
\begin{equation}
V(\Phi)=\frac{1}{2} m^2 |\Phi^2|^2 =\frac{1}{2}m^2(\phi_1^2+\phi_2^2).
\end{equation}

First, consider a small symmetry breaking caused by
\begin{equation}
\Delta V\sim \frac{\Lambda^4}{M^2}\left(\frac{\Phi+\Phi^*}{2}\right)^2,
\end{equation}
where $\Lambda \ll M$ is assumed.
$\phi_2$-oscillation may cause significant particle production
when there is the interaction given by
\begin{equation}
{\cal L}_\mathrm{int}=g(\Phi+\Phi^*)\bar{\psi}\psi,
\end{equation}
which can lead to significant $\psi$-production at the enhanced
symmetric point ($\phi_2\sim0$)~\cite{PR}.
The coefficient of the interaction could be small ($g\sim
\Lambda/M \ll 1$) when it is suppressed by
a cut-off scale.
$\psi$ may decay quickly into radiation since the amplitude of the
oscillation after chaotic inflation is very large~\cite{PR}.

Define $\delta m^2 \equiv \frac{2\Lambda^4}{M^2}$.
If $\delta m^2$ is much smaller than $m^2$, 
the cancellation in Eq.(\ref{spect-1}) is still significant.
On the other hand, it is possible to assume $\delta m^2/m^2\sim O(1)$
(which is still within the conventional set-up of multi-field
inflation) one obtains $P\sim 1$ and $n_s-1\sim
-\frac{1}{N_e}$.
Again, the scenario requires additional inflation
stage~\cite{thermal-Inf}.  

Second, consider the case in which the potential during inflation 
is dominated by a polynomial $V(\Phi)\propto \Phi^p$.
The curvaton can dominate the spectrum, however the spectral index
becomes
\begin{eqnarray}
n_s-1&\simeq& - \frac{M_p^2}{\phi_r^2}\left[p^2-2p(p-1)\right]\nonumber\\
&\simeq& \frac{p-2}{2N_e}.
\end{eqnarray}
The scenario requires $p<2$.

\subsection{Sneutrino inflation}
\label{sn-inf}
It is possible to assume small inflation ``before'' the multi-field
inflation. 
The observed spectrum of the curvaton perturbation exits horizon during
the first inflation.
In that case $\epsilon_H$ is determined by the first inflation and
the cancellation in the spectral index is avoided.
This scenario uses multi-field inflation for the curvaton
inflation~\cite{Infcurv}.

The usual sneutrino inflation~\cite{Ellis:2003sq} uses
$m\sim 10^{13}$ GeV to satisfy the CMB normalization.
When the condition is combined with the gravitino problem, Yukawa coupling of the
first generation sneutrino (single-field inflaton) must satisfy
$(Y_\nu Y_\nu)^{\dagger}_{11}< 10^{-12}$, whilst other Yukawa couplings
will not be so small.
Here $Y_\nu$ is the neutrino Yukawa matrix.

In this section we consider multi-stage inflation, in which three
sneutrinos play crucial role.
We assume that the first single-field inflation is
caused by the third generation sneutrino, and the secondary two-field inflation
is caused by the first and the second generation sneutrinos with the
mass  $M_{1}=M_{2}\equiv \hat{M}$.
We assume $M_{3}>\hat{M}$ for the third generation.

The reheating after two-field inflation is due to the decay of
the second generation sneutrino, which gives the
reheating temperature
\begin{equation}
T_R=\left(\frac{90}{\pi^2g_*}\right)^{1/4}\sqrt{\Gamma_2 M_p},
\end{equation}
where the decay rate is  
\begin{equation}
\Gamma_i\simeq\frac{1}{4\pi}\left(Y_\nu Y_\nu^{\dagger}\right)_{ii}\hat{M}. 
\end{equation}
From Eq.(\ref{zeta-lo}), the curvaton mechanism is significant when
$\Gamma_2\gg \Gamma_1$.
For the two-field sneutrino inflation, which is the secondary inflation
of the above scenario, we find
\begin{equation}
{\cal P}_{\zeta_1}^{1/2} < \frac{1}{3\pi}
\left(\frac{(Y_\nu Y_\nu^\dagger)_{22}}{(Y_\nu Y_\nu^\dagger)_{11}}\right)^{1/4}
\frac{\hat{M}}{M_p}.
\end{equation}
Here the mass of the first (second) neutrino is
\begin{equation}
(m_\nu)_{ii}\simeq (Y_\nu Y_\nu^\dagger)_{ii}\frac{<H_u>^2}{\hat{M}}.
\end{equation}
We thus find for the given neutrino mass $(m_\nu)_{11}$ and $(m_\nu)_{22}$;
\begin{equation}
{\cal P}_{\zeta_1}^{1/2}< \frac{1}{3\pi}
\left(\frac{(m_\nu)_{22}}{(m_\nu)_{11}}\right)^{1/4}
\frac{\hat{M}}{M_p}.
\end{equation}
The reheating temperature after inflation is given by
\begin{equation}
T_R=\left(\frac{45}{8\pi^4g_*}\right)^{1/4}
\frac{\hat{M}}{<H_u>}
\sqrt{(m_{\nu})_{22} M_p},
\end{equation}
while the temperature just after the curvaton decay is
\begin{equation}
T_R'=\left(\frac{45}{8\pi^4g_*}\right)^{1/4}
\frac{\hat{M}}{<H_u>}
\sqrt{(m_{\nu})_{11} M_p}.
\end{equation}
We may write the spectrum ${\cal P}_{\zeta_1}$ using $T_R$ and $T_R'$;
\begin{equation}
{\cal P}_{\zeta_1}^{1/2} < \frac{1}{3\pi}
\left(\frac{T_R}{T_R'}\right)^{1/2}
\frac{\hat{M}}{M_p}.
\end{equation}

When the primary inflation gives the
number of e-foldings $N_1$, the spectral index is
\begin{equation}
n_s-1\simeq -2\epsilon_H\simeq-\frac{1}{N_1}.
\end{equation}
The observation gives $n_s-1= 0.037 \pm 0.014$, which suggests
$20\lesssim N_1\lesssim 40$ for the first inflation.

\subsection{N-flation}
\label{N-fla}

The two-field inflation model considered in this paper is a
simplification of the N-flation model~\cite{N-flation}.
The N-flation has been studied using statistical argument~\cite{CGJ}, 
which helps us understand the results obtained above for the two-field
model.

Assuming (for simplicity) the same potential for all $N_f$ fields, we
find 
\begin{equation}
V(\phi_n)=\sum^{N_f}_{n=1}\frac{1}{2}m^2 \phi_n^2.
\end{equation}
Using the adiabatic field defined by
\begin{equation}
\phi_r^2 \equiv \sum^{N_f}_{n=1}\phi_n^2,
\end{equation}
we find the potential
\begin{equation}
V(\phi_r)=\frac{1}{2}m^2 \phi_r^2.
\end{equation}
If we assume uniform initial condition $\phi_n\simeq \phi_0$,
the model is identical to the two-field model with $\theta
\sim 1/\sqrt{N_f}\ll 1$. 

For the number of e-foldings $N_e\sim 60$, the usual 
curvature perturbation
created at the horizon exit is given by
\begin{equation}
\zeta^\mathrm{inf} =-H_I\left.\frac{\delta \phi_r}{\dot{\phi}_r}\right|_*
=2N_e\left.\frac{\delta \phi_r}{\phi_r}\right|_* .
\end{equation}
where $H_I^2 \equiv \frac{N_f m^2\phi_0^2}{6M_p^2}$ is the Hubble
parameter during the primordial N-flation.

Suppose that the decay rate $\Gamma_{n}$ is uniform {\it except for
a field $\phi_1$}, which has $\Gamma_1\ll \Gamma_n$.
Here the density ratio becomes $r_1^*\simeq \frac{1}{N_f}$.
Repeating the same calculation, we find 
\begin{equation}
\zeta_1\equiv \frac{\delta \rho_1}{3\rho_1}=  
\frac{2}{3} \frac{\delta \phi_1}{\phi_0}\simeq
\frac{2}{3}\sqrt{N_f} \frac{\delta s}{\phi_r}.
\end{equation}
${\cal P}_{\zeta^\mathrm{inf}}^{1/2}\ll {\cal P}_{\zeta_1}^{1/2}$ is
possible when  $N_f \gg N_e^2$. 
This gives the minimum number of the fields that is needed for the
curvaton mechanism and it explains the numerical calculation in
Ref.~\cite{CGJ}.  

In the above scenario, the curvaton is one of the inflaton fields that
are equally participating $1/N_f$ of the inflaton dynamics.

At the end of inflation, the fraction of $\rho_1$ is
\begin{equation}
r_1(t_{end})=\frac{1}{N_f}\ll 1,
\end{equation}
while at the decay of $\phi_1$ it can grow;
\begin{equation}
r_1(t_{decay})=r_1(t_{end})\times
 \left(\frac{\Gamma_n}{\Gamma_1}\right)^{1/2}.
\end{equation}
We need for the curvaton mechanism (i.e, $\zeta_1$-domination)
\begin{equation}
\frac{2}{3}\sqrt{N_f}\frac{{\cal P}_{\delta \phi_1}}{\phi_r}\times \frac{1}{N_f}
\left(\frac{\Gamma_n}{\Gamma_1}\right)^{1/2}>2N_e \frac{{\cal P}_{\delta \phi_r}}{\phi_r},
\end{equation}
which leads to
\begin{equation}
\left(\frac{\Gamma_1}{\Gamma_n}\right)^{1/2}< \frac{1}{3N_e\sqrt{N_f}}.
\end{equation}
Significant non-Gaussianity ($f_{NL}$) requires  
$r(t_{decay})\sim 0.1$, which gives 
\begin{equation}
\left(\frac{\Gamma_1}{\Gamma_n}\right)^{1/2}\sim \frac{10}{N_f}.
\end{equation}
If the distribution is statistical for the decay rate, 
we need $N_f \gg 1$ for the strong suppression ($\Gamma_1/\Gamma_n\ll
1$).

In this section we found that the evolution after inflation may dominate
the curvature perturbation when $N_f$ is large.  
Our result explains the numerical calculation in Ref.~\cite{CGJ}.

\section{Conclusions}

The evolution after multi-field inflation can change the curvature
perturbation. 
In this paper we considered a conventional two-field inflation model and
showed that the curvaton mechanism after multi-field inflation could be significant
when the decay rates are not identical~\footnote{A similar but another story 
has been discussed in Ref.\cite{matsuda-hybrid}.}.
Interestingly, the mechanism works for a complex scalar field
$\Phi\equiv \phi_2+i\phi_1$.

The previous numerical study~\cite{CGJ} showed that $N_f\gg 1$ causes 
significant evolution of the curvature perturbation after inflation as
well as the creation of significant non-Gaussianity.
We showed that the same is true for two-field inflation, in which
 $\theta\ll 1$ is required instead of $N_f\gg 1$.

The source of the curvaton mechanism is the entropy
perturbation generated during multi-field inflation.
Since the uniform density surface of the multi-field potential 
is flat by definition, the perturbation on that surface is inevitable. 

Our results suggest that many-field inflation must be
considered with care.
A large number ($N_f\ge 10^{3}$) can easily explain the required
condition for the curvaton domination.

\section{Acknowledgment}
We thank D.~H.~Lyth for collaboration in the early stage of the paper.
T.M thanks J.~McDonald for many valuable discussions.
S.E. is supported by the Grant-in-Aid for Nagoya University Global COE Program,
"Quest for Fundamental Principles in the Universe: from Particles to the Solar
System and the Cosmos".

\appendix
\section{Calculation details}

\subsection{Evolution of the curvature perturbation}
In this Appendix we show the calculation details of the evolution after
inflation.

We first assume that the potential is quadratic and symmetric during
chaotic inflation.
In our formalism $\zeta^\mathrm{inf}$ is defined  at
the end of inflation.
The entropy perturbation is realized by $\delta \theta$, which is the
perturbation of the angle $\theta$ in Fig.\ref{fig:equalmass}.

The spectrum of the entropy perturbation during inflation is ${\cal
P}_{\delta s_*} \simeq (H_{*}/2\pi)^2$.
The entropy perturbation causes the fraction perturbation between densities.
Using $\delta\theta$, the densities of the components and the 
isocurvature perturbations at the end of inflation are given by 
\begin{eqnarray}
\bar{\rho}_{1,\mathrm{end}}&=&\frac{1}{2}m^2|\phi_r^\mathrm{end}|^{2}\sin^2 \bar{\theta} \simeq
\frac{m^2M_p^2}{2}\sin^2 \bar{\theta}\\
&&\delta \rho_{1,\mathrm{end}}^\mathrm{iso}\simeq m^2 M_p^2(\sin\bar{\theta} \cos\bar{\theta} )
\delta\theta
,\\
\bar{\rho}_{2,\mathrm{end}}&=&\frac{1}{2}m^2|\phi_r^\mathrm{end}|^{2}\cos^2 \bar{\theta} \simeq 
\frac{m^2M_p^2}{2}\cos^2 \bar{\theta}\\
&&\delta \rho_{1,\mathrm{end}}^\mathrm{iso}+\delta \rho_{2,\mathrm{end}}^\mathrm{iso}=0.
\end{eqnarray}
We find at the end of inflation:
\begin{eqnarray}
f_1&\equiv& \frac{\bar{\rho}_1}{\bar{\rho}_1+\bar{\rho}_2}= \sin^2\bar{\theta},\\
\delta f_1&\simeq& \frac{\partial
f_1}{\partial \theta} \delta \theta
=2[\sin\bar{\theta}\cos\bar{\theta}]\delta \theta\nonumber\\
&=&[\sin2\bar{\theta}] \delta \theta.
\end{eqnarray}
The expansion with respect to $\delta \theta$
makes no sense when $\delta \theta/\sin\theta\ge 1$ or $\delta
\theta/\cos\theta \ge 1$~\cite{Lyth-ngaus}. 
We are excluding those regions.

Creation of the curvature perturbation after inflation requires 
the decay rate $\Gamma_1 \ll \Gamma_2$.
In the phase (A) we find 
\begin{eqnarray}
\zeta^\mathrm{iso}_{1A}&\simeq&\frac{2}{3}
\frac{\cos\bar{\theta}}{\sin\bar{\theta}}\delta 
 \theta\\
\zeta^\mathrm{iso}_{2A}&\simeq&-\frac{1}{2}\frac{\sin\bar{\theta}}{\cos\bar{\theta}}\delta
 \theta.
\end{eqnarray}
Using Eq.(\ref{deltaN-1}), we find
\begin{eqnarray}
\label{zeta-lo}
\zeta^\mathrm{fin}
&=&
\left[\frac{2}{3}r_{1-}\frac{\cos\bar{\theta}}{\sin\bar{\theta}}-
\frac{1}{2}(1-r_{1-})
\frac{\sin\bar{\theta}}{\cos\bar{\theta}} \right]\delta \theta
+\zeta^\mathrm{inf}\nonumber\\
&=&\left[\frac{4r_{1-}\cos^2\bar{\theta}-3(1-r_{1-})\sin^2\bar{\theta}}
{6\sin\bar{\theta}\cos\bar{\theta}}
 \right]\delta \theta\nonumber\\
&&+\zeta^\mathrm{inf},
\end{eqnarray}
where $r_{1-}$ denotes the value of $r_{1A}$ evaluated just before the
end of the phase (A).

The evolution is
\begin{eqnarray}
\bar{\rho}_{1-}&=&\left[\frac{m^2 M_p^2}{2}\sin^2\bar{\theta}\right]\times
\left(\frac{a_\mathrm{d1}}{a_\mathrm{end}}\right)^{-3}
\nonumber\\
\bar{\rho}_{2-}&=&\left[\frac{m^2 M_p^2}{2}\cos^2\bar{\theta}\right]\times
\left(\frac{a_\mathrm{d2}}{a_\mathrm{end}}\right)^{-3}
\left(\frac{a_\mathrm{d1}}{a_\mathrm{d2}}\right)^{-4},
\end{eqnarray}
which leads to the ratio
\begin{eqnarray}
\frac{\bar{\rho}_{2-}}{\bar{\rho}_{1-}}&=&\frac{\cos^2\bar{\theta}}{\sin^2\bar{\theta}}
\left(\frac{a_{d2}}{a_{d1}}\right).
\end{eqnarray}
Therefore, in the radiation dominated Universe we find
\begin{eqnarray}
\label{r1-}
r_{1-}&=&\frac{3\rho_{1-}}{3\rho_{1-}+4\rho_{2-}}\nonumber\\
&=&\frac{3\sin^2\bar{\theta}}{3\sin^2\bar{\theta} + 4\cos^2\bar{\theta}
\sqrt{\Gamma_1/\Gamma_2}}.
\end{eqnarray}
Domination by the curvaton density ($r_{1-}\sim1$) requires 
$\sqrt{\Gamma_1/\Gamma_2}\le \tan^2\bar{\theta}$.

The CMB spectrum requires
${\cal P}_{\zeta^\mathrm{fin}}\simeq (5\times 10^{-5})^2$~\cite{WMAP7}.
The requirement is trivial when
$\zeta^\mathrm{fin}\simeq\zeta^\mathrm{inf}$,~\footnote{Note
however the non-Gaussianity is 
not trivial because the curvaton perturbation may still dominate the
second-order perturbation~\cite{Lyth-adi0}.} while in the
opposite case $\zeta^\mathrm{fin}>\zeta^\mathrm{inf}$, in which the
curvaton mechanism dominates,
we need the condition
\begin{equation}
\label{adi-iso-con}
\left[\frac{2}{3}r_{1-}\frac{\cos\bar{\theta}}{\sin\bar{\theta}}-
\frac{1}{2}(1-r_{1-})
\frac{\sin\bar{\theta}}{\cos\bar{\theta}} \right]\delta \theta >
\frac{\delta \phi_{r*}}{\eta \phi_{r*}}.
\end{equation}
Solving Eq.(\ref{adi-iso-con}) for $r_{1-}$ and using Eq.(\ref{r1-}), we
find 
\begin{equation}
\label{gamma12}
\sqrt{\frac{\Gamma_1}{\Gamma_2}}<
\frac{2\eta\tan\bar{\theta}-3\tan^2\bar{\theta}}{4+2\eta\tan\bar{\theta}}<1.
\end{equation}
This equation also shows that $2\eta-3\tan\bar{\theta}>0$, which gives
\begin{equation}
\label{justif-theta}
\tan\bar{\theta}<\frac{2}{3}\eta.
\end{equation}

The CMB observation gives the normalization
\begin{eqnarray}
\label{spectrum0}
\left[\frac{2}{3}r_{1-}\frac{\cos\bar{\theta}}{\sin\bar{\theta}}-
\frac{1}{2}(1-r_{1-})
\frac{\sin\bar{\theta}}{\cos\bar{\theta}} \right]
{\cal P}^{1/2}_{\delta \theta}
\simeq 5\times 10^{-5}.
\end{eqnarray}

Defining $k\equiv \frac{{\cal P}^{1/2}_{\delta\theta}}{5\times
 10^{-5}}$ and $y\equiv
\sqrt{\Gamma_1/\Gamma_2}$, we can solve Eq.(\ref{spectrum0}) for $y$ and find
\begin{eqnarray}
\label{zeta-sim}
y
&=& \frac{2k -3\tan\bar{\theta}}
{2k+4\tan^{-1}\bar{\theta}}\nonumber\\
&\simeq & \frac{k\bar{\theta}}{2}.
\end{eqnarray}
To avoid $y<0$, we need the condition
\begin{equation}
 \frac{3}{2}\tan\bar{\theta}<k.
\end{equation}

The perturbations can be expanded up to second order.
We find
\begin{eqnarray}
\label{zeta12ndorder}
\zeta^\mathrm{iso}_1&\simeq&\frac{2}{3}
\left[\cos\bar{\theta} \left(\frac{\delta \theta}{\sin\bar{\theta}}\right)
+\frac{1}{2}\cos2\bar{\theta} \left(\frac{\delta \theta}{\sin\bar{\theta}}\right)^2
\right]
\\
\zeta^\mathrm{iso}_2&\simeq&-\frac{1}{2}
\left[\sin\bar{\theta} \left(\frac{\delta \theta}{\cos\bar{\theta}}\right)
+\frac{1}{2}\cos2\bar{\theta} \left(\frac{\delta \theta}{\cos\bar{\theta}}\right)^2
\right].
\end{eqnarray}
Using Eq.(\ref{deltanbc}), the final curvature perturbation after the
decay is
\begin{eqnarray}
\label{zeta-losecond}
\zeta^\mathrm{fin}
&=&
\left[\frac{2}{3}r_{1-}\frac{\cos\bar{\theta}}{\sin\bar{\theta}}-
\frac{1}{2}(1-r_{1-})
\frac{\sin\bar{\theta}}{\cos\bar{\theta}} \right]\delta \theta\nonumber\\
&&
+\left[\frac{1}{3}r_{1-}\frac{\cos2\bar{\theta}}{\sin^2\bar{\theta}}
-\frac{1}{4}(1-r_{1-})
\frac{\cos2\bar{\theta}}{\cos^2\bar{\theta}} \right](\delta \theta)^2\nonumber\\
&&+\zeta^\mathrm{inf}\nonumber\\
&=&\left[\frac{4r_{1-}\cos^2\bar{\theta}-3(1-r_{1-})\sin^2\bar{\theta}}
{3\sin2\bar{\theta}}
 \right]\delta \theta\nonumber\\
&&+\frac{\cos2\bar{\theta}}{3\sin^2 2\bar{\theta}}
\left[4r_{1-}\cos^2\bar{\theta}-3(1-r_{1-})\sin^2\bar{\theta}
\right](\delta
\theta)^2\nonumber\\ 
&&+\zeta^\mathrm{inf}\nonumber\\
&=&
\frac{4r_{1-}\cos^2\bar{\theta}-3(1-r_{1-})\sin^2\bar{\theta}}
{3\sin2\bar{\theta}}\nonumber\\
&&\times\left[\delta \theta + \frac{\cos2\bar{\theta}}{\sin2\bar{\theta}}(\delta \theta)^2
	\right]
+\zeta^\mathrm{inf}.
\end{eqnarray}

When the curvaton perturbation dominates ($\theta \ll 1$), 
the non-Gaussianity of the spectrum is measured by
\begin{eqnarray}
\label{fnl-stdd}
f_{NL}&\simeq&\frac{5\cos2\bar{\theta}}
{4r_{1-}\cos^2\bar{\theta}-3(1-r_{1-})\sin^2\bar{\theta}}.
\end{eqnarray}

Using Eq.(\ref{r1-}), we can substitute $r_{1-}$ in Eq.(\ref{fnl-stdd}). 
Then solving the equation for $y$, we find
\begin{eqnarray}
\label{gammma12a}
y&\simeq&\frac{3}{4}\tan^2\bar{\theta}\left[
\frac{4}{5}\frac{\cos^2\bar{\theta}}{\cos2\bar{\theta}}f_{NL}-1\right].
\end{eqnarray}
Barring cancellation, the above equation gives a simplified formula
\begin{eqnarray}
\label{appy}
y&\simeq&
\frac{3}{5}f_{NL}\bar{\theta}^2.
\end{eqnarray}
Being combined with Eq.(\ref{zeta-sim}), which has been obtained using
the CMB normalization, we find
\begin{eqnarray}
k& \simeq& \frac{6}{5}f_{NL} \bar{\theta}.
\end{eqnarray}
We thus find (from $f_{NL}$ and CMB using the definition of $k$)
\begin{eqnarray}
\frac{{\cal P}^{1/2}_{\delta\theta}}{\bar{\theta}}&\simeq&6\times 10^{-5}
 \times f_{NL}
\end{eqnarray}
or equivalently 
\begin{eqnarray}
\label{HIeq}
H_I
&\simeq&
6\times 10^{-3}\times
f_{NL}\bar{\theta}M_p.
\end{eqnarray}
Solving the equation for $\bar{\theta}$, it gives
\begin{eqnarray}
\bar{\theta}&\simeq & \frac{1}{6f_{NL}}
\left[\frac{H_I}{M_p}\times 10^3\right].
\end{eqnarray}

Using $H_I$ in Eq.(\ref{HIeq}) and calculating the tensor to
scalar ratio $r_g$, we find~\cite{Lyth-bound} 
\begin{equation}
r_g\simeq f_{NL}^2 \bar{\theta}^2\times 10^4.
\end{equation}

Considering the natural bound $\Gamma_2<H_I$ and
$\Gamma_1>H_\mathrm{nuc}$, where $H_\mathrm{nuc}$ is the Hubble
parameter at the time of the nucleosynthesis,
Eq.(\ref{gammma12a}) gives the lower bound for $\bar{\theta}$;
\begin{equation}
\bar{\theta}> \left(\frac{H_\mathrm{nuc}}{H_I}\right)^{1/4}.
\end{equation}
Besides the above condition, we have another condition coming from
$\bar{\theta}>\delta \theta$.
Since we are assuming quadratic potential in the
trans-Planckian, we have $\delta \theta =\delta s/\phi_{r*}$
 and $\phi_{r*}=2\sqrt{N_e}M_p$.
Then $\bar{\theta}>\delta \theta$ leads to 
\begin{equation}
\bar{\theta} >0.01 \frac{H_I}{M_p}.
\end{equation}

\end{document}